# EFL Students' Attitudes and Contradictions in a Machine-in-the-loop Activity System

David James Woo [a, *], Hengky Susanto [b] and Kai Guo [c]

[a] Precious Blood Secondary School, Hong Kong, China

[b] Education University of Hong Kong, Hong Kong, China

[c] Faculty of Education, The University of Hong Kong, Hong Kong, China

[*] Corresponding author

- Postal address: Precious Blood Secondary School, 338 San Ha Street, Chai Wan, Hong Kong, China
- Email address: net_david@pbss.hk
- Phone: +852 2570 4172

## Disclosure Statement

The authors report there are no competing interests to declare.

## Data Availability Statement

The data that support the findings of this study are available from the corresponding author, David James Woo, upon reasonable request.

## Biographical Note

***David James Woo*** is a secondary school teacher. His research interests are in artificial intelligence, natural language processing, digital literacy, and educational technology innovations. ORCID: https://orcid.org/0000-0003-4417-3686

***Hengky Susanto*** received his BS, MS and PhD degree in computer science from the University of Massachusetts system. He was a postdoctoral research fellow at University of Massachusetts Lowell and Hong Kong University of Science and Technology. He was also senior researcher at Huawei Future Network Theory Lab. Currently, he is a principal researcher in a startup mode research laboratory and a lecturer at Education University of Hong Kong. His research interests include applied AI (computer vision and NLP) to solve complex social problems, smart city, and computer networking (e.g., datacenter network, congestion control, etc.).

***Kai Guo*** is a Ph.D. candidate in the Faculty of Education at the University of Hong Kong. His research focuses on second language writing, technology-enhanced learning, and artificial intelligence in education. His recent publications have appeared in international peer-reviewed journals such as *Computers & Education*, *Education and Information Technologies*, *Interactive Learning Environments*, and *Journal of Educational Computing Research*. ORCID: https://orcid.org/0000-0001-9699-7527

# EFL Students' Attitudes and Contradictions in a Machine-in-the-loop Activity System

## Abstract

This study applies Activity Theory and investigates the attitudes and contradictions of 67 English as a foreign language (EFL) students from four Hong Kong secondary schools towards machine-in-the-loop writing, where artificial intelligence (AI) suggests ideas during composition. Students answered an open-ended question about their feelings on writing with AI. Results revealed mostly positive attitudes, with some negative or mixed feelings. From a thematic analysis, contradictions or points of tension between students and AI stemmed from AI inadequacies, students' balancing enthusiasm with preference, and their striving for language autonomy. The research highlights the benefits and challenges of implementing machine-in-the-loop writing in EFL classrooms, suggesting educators align activity goals with students' values, language abilities, and AI capabilities to enhance students' activity systems.

## Introduction

The capability of artificial intelligence (AI) to generate natural language (NLG) text indistinguishable from human-written text has captivated educational researchers and practitioners (Hwang & Chen, 2023). AI's capability for NLG should not be used to replace human effort but can potentially augment human effort so that AI remains under human control and improves human activity (Yang et al., 2023). For the activity of writing, efforts have been made to conceptualize human-AI collaboration as writing with a machine-in-the-loop (Clark et al., 2018). In this framework, the human is the central actor and AI plays a supporting role so that when a person needs help, such as when searching for ideas, the AI can suggest ideas and the person has full agency to decide what to do with the AI's ideas, if anything (Author, 2023). On the other hand, there is evidence that AI can play a larger role in machine-in-the-loop writing, being perceived more as an active writer (Yang et al., 2022). Besides, a person's approach and motivation can influence the human-AI dynamic when writing with a machine-in-the-loop (Singh et al., 2022).

In the context of the English as a foreign language (EFL) classroom, students may experience different attitudes and expectations when writing with a machine-in-the-loop. This affective dimension of machine-in-the-loop writing remains relatively unexplored, but has potential implications for student engagement, motivation and learning outcomes. This study explores students' attitudes and tension points when writing with a machine-in-the-loop, applying Activity Theory (AT) and framing AI's role as that of an NLG tool.

## Literature Review

### Applying AI Tools in EFL Writing Classrooms

EFL students often encounter challenges when writing in a non-native language, such as struggling with grammar, vocabulary, syntax (De Wilde, 2023; Vasylets & Marín, 2021), lacking confidence in their writing abilities (Sun & Wang, 2020; Zotzmann & Sheldrake, 2021), and encountering difficulties in generating ideas during the writing process (Crossley et al., 2016; Hayes & Flower, 2016). Previous research has investigated and advocated different pedagogical approaches to enhancing EFL students' writing skills. These include process models and genre-based pedagogies (Hyland, 2003). Importantly, collaborative writing has been extensively implemented and has shown favorable effects on students' writing performance (Elabdali, 2021) and attitudes (Yesuf & Anshu, 2022). However, in practice, it can be challenging for students to identify an ideal writing partner.

Some scholars propose the use of AI applications as a learning companion to assist EFL students in their writing (Author, 2022; Su et al., 2023). There has been progress in the development and implementation of these applications, largely at the university level. For example, Zhang et al. (2023) developed a chatbot with which 15 Chinese EFL undergraduate and postgraduate students trained on logical fallacy in argumentative writing. Some applications show benefits for student learning: Annamalai et al. (2023) explored EFL university students' learning with chatbots and found chatbots supported these learners' psychological needs. On the other hand, these applications may raise new complications. For instance, Yan (2023) found Chinese undergraduate students in an EFL major program expressed concerns about the possible threat of ChatGPT to educational equity and academic honesty. In contrast, there is a noticeable gap in the literature regarding the usage of AI applications by EFL learners at primary and secondary levels during lessons (Jeon, 2022; Ji et al., 2023), especially these younger EFL students' experiences and perspectives when writing with AI.

The language models that power an AI's NLG capability are advancing rapidly. The advancements include breakthroughs in neural network architecture (Vaswani et al., 2017), training (Hu et al., 2021), compression (Xiao et al., 2022) and fine-tuning (Wang et al., 2023). As a result, an increasingly diverse range of language models can now produce coherent text indistinguishable from human writing and power AI writing applications (Brown et al., 2020). These range from large commercial language models such as those in ChatGPT to open-source language models that are free to use and access on the Hugging Face machine learning repository (HuggingFaceH4, 2023). The proliferation of language models on AI writing applications in the EFL classroom are manifold but largely unexplored. Language models will perform language tasks in different ways and to different degrees of proficiency. Naturally these performance differences may result in different positive and negative impacts on EFL students in the writing classroom.

**Activity Theory**

AT, proposed by Vygotsky (1978) and Leontyev (1981), provides a useful framework for analyzing human activity in sociocultural contexts. According to Vygotsky's (1978), by human activity a person interacts with an object by utilizing tools that have been developed within a specific sociocultural context, leading to both intended and unintended consequences. Leontyev (1981) further elaborated that such mediated activity is embedded in a social environment or community with an evolving division of labor. Engeström (1987) expanded on these elements and their structure (see Figure 1) by asserting that human activity takes place within a community comprising a division of labor and rules that dictate how subjects use tools to achieve objects. These rules can include instructions, assumptions, and established practices.

Figure 1.

Activity system

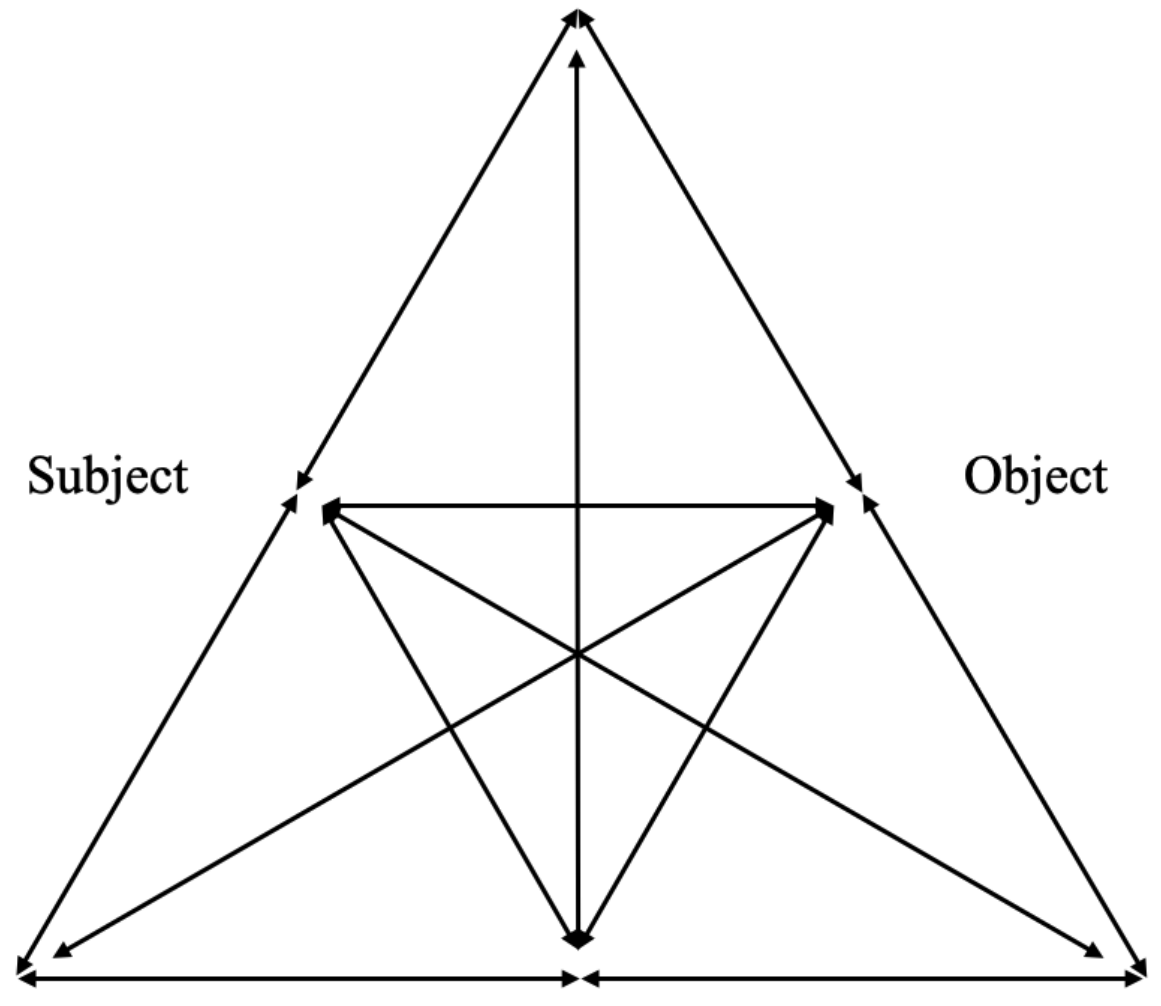

An essential mechanism to drive change and development of an activity system is contradiction (Engeström, 2001). According to Engeström (2001), contradiction refers to tension points within and between activity systems but not necessarily problems or conflicts. Engeström (1987) had defined four types of contradiction: *a primary contradiction* exists within an element in an activity system; *a secondary contradiction* exists between elements in an activity system, for instance, between a tool and an object; *a tertiary contradiction* exists between an old and a new activity system; and *a quaternary contradiction* exists between an activity system and neighboring systems. Attempts to resolve contradictions can lead to innovations and improvement of activity systems.

*Attitude* refers to a person's inclination to respond positively or negatively to a certain idea, object, person, or situation, based on their perceptions and evaluations (Vargas-Sánchez et al., 2016). Although AT does not explicitly address attitude, it is integral to human behavior and thus the subject in an activity system. We conceptualize that attitude can influence the subject's engagement in an activity system and other system elements can change or disturb the subject's attitude. In this way, attitude, especially a negative attitude, may indicate contradiction within an activity system. By considering the role of attitude within the context of human activity, we can possibly improve activity systems that lead to better learning.

We use AT to conceptualize as a mediated activity system the context of writing with a machine-in-the-loop in an EFL classroom. At its core, an EFL student is the human subject that interacts with AI, which is an NLG tool, to realize the object of a writing composition. The intended outcome is to complete the writing composition. Although each student shares the same community, NLG tools, rules, etc., we conceptualize that students' attitudes, the attributes of the writing compositions and the division of labor or responsibilities between the student and the NLG tool are different. This is not least because an EFL student's attitude can be influenced by the student's desires on the object and an NLG tool's output. In this way, each student has their own activity system.

We focus our observations on the triad of subject, object, and tools in a machine-in-the-loop activity system (see Figure 2). We expect to observe similarities and differences in students' activity systems, including patterns in student attitudes and with some students, the emergence of a contradiction or point of tension between the student, the NLG tool and the writing composition. We are interested in exploring this secondary contradiction, identifying its attributes so as to possibly resolve them and to improve students' activity systems.

Figure 2.

Integrating attitude into the triad of subject, object, and tools in a machine-in-the-loop activity system

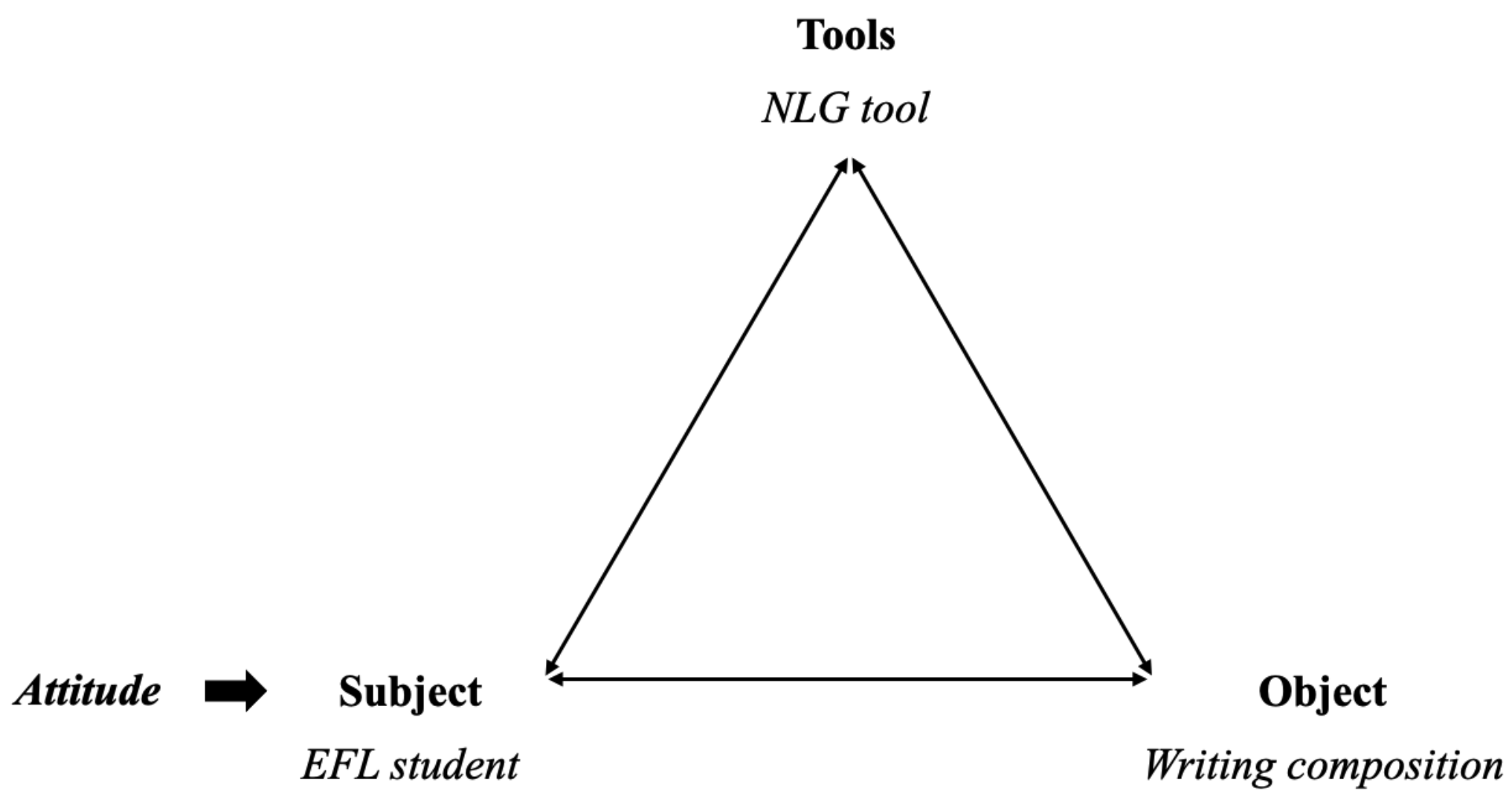

## Research Questions

1. What attitudes do EFL students report when participating in a machine-in-the-loop activity system?
2. What attributes of tension emerge between a student, an NLG tool and a writing composition in a machine-in-the-loop activity system?

# Material and Methods

## Participants

Participants were 67 EFL students from four Hong Kong secondary schools which recruit students at different levels of academic achievement based on the students of each school's university entrance exam results. Each school enrolled its students in a workshop where they received instruction on machine-in-the-loop writing and were given 45 minutes to practice writing a short story of no more than 500 words with a machine-in-the-loop. Students should use their own words and words generated from NLG tools. Figure 3 shows a story written using a student’s own words in italics and red and words generated from NLG tools in non-italicized, black text. Each school’s workshop was delivered by the first author, between December 2022 and April 2023.

Figure 3.

A story written using a student’s own words and words generated from an NLG tool

*On a chilly winter night, Maddison enters her bedroom at midnight and falls upon what she thinks is a pen. She picks it up and it turns out to be… a crystal? She said ,* “woah! This crystal is ginormous ! where did this crystal come from?”

*Soon, the crystal* starts glowing a bright purple*, a bright light flashes her eyes and teleports her to an unknown realm.* Then the light turns into an orb of energy. The orb introduces itself as the crystal king, he wants to hire maddie as *his right hand to take over Lego mania.* The orb gives maddison a choice: join him. After a moment, she agrees. Then, The crystal king also gives her a gift, *a crystal.*

*After Maddison awakens, the crystal tells her to go to her* old boarding school and plant the crystal in the *bonsai plant* on the front counter. When she woke up, she then witnessed the *crystal* grow bigger and bigger eventually turning the school into a fortress.

*Soon, six colour coded ninja come, the green one steps forward and asks. ‘Harumi how are you alive!? I saw you get crushed in a crumbling building. He* said that there was so much rubble that he couldn’t find you after days and nights. *Maddison then answers with “what are you talking about? I’m Maddison, who is harumi?’* Everyone had a shocked look on their face *by that statement.* You continue saying *‘found this crystal in my room and this orb named the crystal king came, hired me to help him take over somewhere called Lego mania’.* Wait, so you aren’t the hired villain for *Harumi?* Asked *the blue ninja*. *Maddison* wonders if she looks that much different with eye makeup. ‘Heavens no! I’m an ordinary school *girl* from *Chicago’ replied Maddison*.

*With that, the crystal king snapped back with* ‘I don’t hire children as villians, I’m going to send you back’ *A spark of light appeared from the crystal king's hand and Maddison was transported from Lego Mania to her bedroom.*

*Maddison* jumps into her bed with a feeling of deja vu, not knowing what just happened. She eventually fell asleep.

Students had composed the NLG tools that they used for the machine-in-the-loop writing task. They composed the tools using Python programming language and open-source language models on Hugging Face (https://huggingface.co/). The language models used included GPT2, Bloom560M and GPT-J-6B. Although there was variation in the interface of NLG tools, Figure 4 shows a typical interface of a student's NLG tool where a student can input a prompt in the text box located at the top and the tool generates its output text at the bottom.

Figure 4.

Interface of a student's NLG tool on Hugging Face (identifiers removed)

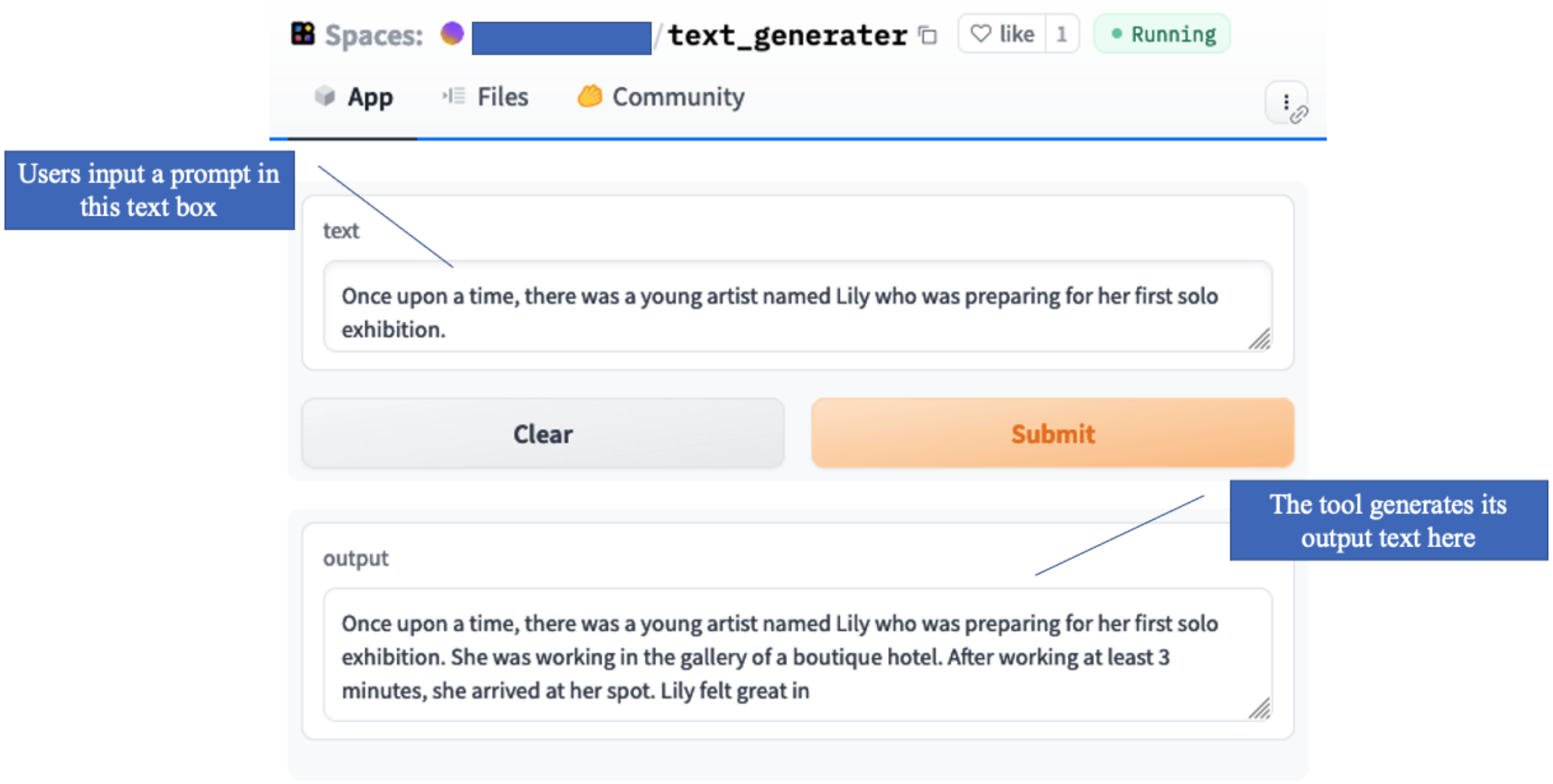

## Data Collection

At the end of the workshop, students answered the open-ended question: "How do you feel writing a short story using your own words and a machine's words?" Students provided their answers on a Google form. We anonymized data for analysis, removing students' names and their school information.

## Data Analysis

To answer the first research question, we manually coded students' answers, as this analysis method has been shown to lead to more valid coding performance of attitude than crowd coding, using dictionaries or machine learning (van Atteveldt et al., 2021). First, we read through all answers to get a sense of the range of attitude. We then open-coded (Saldana, 2012) students' answers to the question item, categorizing any student answer as either negative attitude or positive attitude. We iteratively developed codes for negative attitude and positive attitude and their definitions. We present descriptive statistics of our analysis, and representative excerpts of the codes.

To answer the second research question, we performed a thematic analysis (Braun & Clarke, 2006) on students' answers to the question item. To start, we open-coded students' answers, developing and applying codes for influences on student attitude regardless of the specific attitude. For insight into contradictions within a student's activity system, we examined the influence codes associated with students' negative attitudes. We performed cross-tabulation, re-reading students' responses that for the first research question had been coded for negative attitude; we looked for common influences in these answers and grouped codes for influences into broader categories or themes that capture the essence of the data. Otherwise, a code for influence stands alone as its own theme. We report each theme of tension between the human subject, the NLG tool and the story object in an activity system.

To enhance the trustworthiness of the open-coding, we used multiple coders so that the first author initially developed the sentiment and influence codes and coded the data independently; the second author reviewed the codes and coded the data independently; and the third author reviewed the codes and the other authors' coding, the authors discussing and resolving any code or coding discrepancies. Furthermore, we compiled a codebook and calculated the basic proportion of agreement (DeCuir-Gunby et al., 2010) for the first and second authors' coding (See Appendix 1 for the codebook and Appendix 2 for the coding application). The authors arrived at one-hundred-percent agreement for coding positive and negative attitudes. The three authors worked together to develop the themes.

# Results

## Attitudes

From the 67 student responses, we threw out responses from three students: Students 15 and 17 provided no response; and Student 35 wrote ‘255’, which was irrelevant. .

Of the remaining 64 student responses, we found 54 students showed a positive attitude in their responses. Students that expressed positive attitudes used words such as 'amazing' (Students 5, 23, 39, and 67), 'cool' (Students 8, 57 and 58), 'enjoyable' (Student 9), ‘excited’ (Student 7), ‘exciting’ (Student 30), ‘good’ (Students 13, 22, 23, 24, 25, 26, 43, 50, 57 and 64) and 'useful' (Students 30, 46, 49) in their responses.

Sixteen students showed a negative attitude in their responses. Students that expressed negative attitudes used words such as 'AI is so bad' (Student 21), 'complicated' (Students 4 and 33), 'difficult' (Students 44, 48 and 59), 'no sense' (Student 52) and ‘tired’ (Student 6) in their responses.

Eight students expressed both positive and negative attitudes, that is, mixed feelings in their responses. Thus, 46 students showed exclusively a positive attitude in their response and eight students negative.

Two students in the study did not express any particular attitude towards writing with a machine-in-the-loop. Their responses were, 'I don't have any feeling' (Student 38) and 'No' (Student 37).

## Attributes of Tension

Our thematic analysis revealed three themes from students’ responses that were associated with negative attitude or mixed feelings attitude. The three themes encapsulate attributes of tension between a student, an NLG tool and a story. Since the themes represent an integration of codes across many student answers, the themes are not mutually exclusive and some student responses reflect multiple themes. In this way, quotes were chosen to be representative examples of each theme but the selected quotes may be attributed to more than one theme. Student numbers and attitudes are given in parentheses.

*AI's inadequacies (code: AI limitations)*

Some students reported the negative attributes of NLG tools, pointing out the limitations of using text from NLG tools in writing a story. These include the peculiarities of NLG tools' output generation and constraints on NLG tools' reasoning capabilities:

- I'm sad to say that the text generators don't have much logic. But it was still very fun to try taking pieces of its responses for my story. (Student 20; mixed feelings)
- I think it's kinda useful but sometimes it doesn't writes the way I wanted it to be (Student 46; mixed feelings)
- Some machine's word is i never see it , the words is difficult but it can let my story more interesting；if i use my own words maybe i would not always can use the correct grammar because I 'm not carefully.（Student 48; mixed feelings)

*Balancing enthusiasm with preference (codes: convenience; interesting)*

Some students expressed curiosity, excitement or interest after writing with a machine-in-the-loop. Others appreciated the convenience and ease of writing with NLG tools. At the same time, these students reported the trade-off between their positive attitudes and appreciation with with potential pitfalls or deviation from their own preferences and needs:

- Nice and fun. I can see what the AI thinks and how differently and how interesting the AI can think. It could be weird and funny. But it was such a nice interesting experience and i learnt honestly a lot (Student 32; mixed feelings)

- Interesting and challenging (Student 47; mixed feelings)
- Sometimes quite confusing but also quite convenient (Student 53; mixed feelings)

*Struggling for language autonomy (codes: language issues; own words)*

Some students referred to words, sentences and other language support provided by NLG tools but expressed a preference for their own words or sentences in story writing. In other words, students desired to express themselves but grappled with the NLG tools' syntactical choices:

- It's more easy just using own words (Student 14; negative)
- Using my own words is a little harder but faster and using a machine might takes some time but it's funny. (Student 29; mixed feelings)
- I feel more powerful than the ai because the ai doesn't understand simple sentences and makes it too complicated. (Student 33; mixed feelings)
- I like to use my own words more (Student 66; negative)

## Discussion

The EFL students showed varying attitudes when writing in a machine-in-the-loop activity system. The majority of students reported a positive attitude towards writing with a machine-in-the-loop. This finding aligns with previous research demonstrating that collaborative writing favorably impacts EFL students' attitudes (Yesuf & Anshu, 2022) and EFL students' positive perceptions after using NLG tools in lessons (Jeon, 2022). However, a subset of students expressed a negative attitude or mixed feelings, indicating that points of tension may exist when students write with a machine-in-the loop.

We found students' negative attitudes could be encapsulated by three themes, or attributes of tension between the student, the NLG tools and the story. First, AI limitations were a salient concern for students and the specific NLG tool limitations that our study finds associated with students' negative attitudes, such as NLG tool's poor reasoning ability, incoherence and inability to generate students' desired output are different from those limitations and other factors that are associated with students' negative attitudes in previous studies (Annamalai et al., 2023; Zhang et al., 2023). Our student's specific limitations might be attributed to the smaller, open-source language models used in the NLG tools.

Second, some students in the present study appear aware of both the benefits and constraints of writing with a machine-in-the-loop. Although they were a minority within the study's student population, they highlight another human-AI collaborative context where educators might turn tensions between students and AI into opportunities for learning, instead of blindly assuming that the mere presence of AI guarantees better support for students (Moore et al., 2022). In this way, we do not go as far as other scholars to generalize that the affordances of NLG tools as learning tools mitigate any negative perceptions of the tools' technical limitations (Jeon, 2022).

Third, some students struggled for language autonomy and this indicates these students might at best benefit from the original conceptualization of writing with a machine-in-the-loop (Clark et al., 2018), where AI plays a minor role and the student remains the central actor. These students' struggles can evidence why AI should not play a larger role in machine-in-the-loop writing (Yang et al., 2022), in the context of the EFL classroom.

In sum, the different attitudes and attributes of tension found in our study indicate that students had a range of experiences when writing with a machine-in-the-loop. We suggest these heterogeneous experiences may have been influenced by individual factors such as the different expectations, language proficiency and views on AI that students brought into the activity system.

**Implications**

This study contributes to our understanding of students as subjects in a machine-in-the-loop activity system. The themes from this study illuminate some of the complex dynamics that students grapple with and that emerge within machine-in-the-loop activity systems. There are potential benefits and challenges to implementing machine-in-the-loop writing in the EFL classroom: if an educator were to implement, the tensions from our study point to a need for aligning activity goals with students' own values for story writing, language capabilities and NLG tool capabilities.

In this way, we recommend educators understand students' attitudes so as to support personalized learning (Hwang & Chen, 2023) in a machine-in-the-loop activity system. Particularly by identifying negative attitudes and the influences on attitudes, according to AT, teachers and students may play a role not least as a community to address the contradictions in a student's activity system, improving that system and leading to student learning. For example, teachers can manage students' expectations of NLG tools' capabilities. Additionally, teachers may adjust a student's rules within the activity system, such as reducing the expected number of words from NLG tools in the writing composition, reducing the number of instances to use NLG tools, and narrowing the variety of ideas from NLG tools for the composition. A teacher may select a commercially available NLG tool such as ChatGPT or, if finances are a constraint, from an increasing number of sophisticated, free-to-use NLG tools on Hugging Face Spaces. Not least with improved NLG tools and well-designed writing activities, machine-in-the-loop activity systems could provide more enjoyable and useful experiences for more EFL students.

Nonetheless, we also acknowledge that a language teacher implementing these practical guidelines in the classroom demands stronger orchestration, such that this AI-integration may enhance student learning but not decrease a teacher's workload (Ji et al., 2023).

**Limitations and Future Research**

This study examined only one secondary contradiction between the human-subject, the NLG tool and the story object. Theoretically there are more possible secondary contradictions beyond this triad of elements in an activity system, involving other elements in the activity system. Besides, future research should investigate strategies for mitigating points of tension. Investigating such contradictions and how to resolve them leads to innovation and improvement in an activity system.

This study examined student attitudes from only one writing task with a machine-in-the-loop. Longitudinal studies could shed light on the development of students' attitudes and experiences over time, providing valuable information on the effectiveness of EFL writing with a machine-in-the-loop. Exploring individual factors that contribute to the different attitudes observed in the study can facilitate educators delivering more practical and personalized support for students when writing with a machine-in-the-loop.

This study examined NLG tools composed from smaller, open-source language models. Further research with NLG tools should include the larger, more sophisticated open-source language models that have been released rapidly since the time of the present study. These more recent and advanced models may influence students' attitudes differently and influence different attributes of tension between the student, the NLG tool and the writing composition.

Word count: 4,917 words

(including abstract and references; excluding appendices)

Appendix 1.

Coding scheme

| Code name | Definition | Example response | Basic proportion of agreement (Total no. of agreements) |
|---|---|---|---|
| AI limitations | A student describes any negative attributes of AI, machine or text generator; pointing out the limitations or drawbacks of using AI-generated text in writing a story. The student's answer is more substantial than, "AI is so bad." | *“Using my own words is a little harder but faster and using a machine might takes some time but it’s funny.”* | 1 (5) |
| Convenience | A student refers to convenience or ease (e.g. easy; easily) of writing with the machine’s words | *“It's not bad. It s make writing more easy”* | 1 (5) |
| Fun | A student explicitly expresses amusement, entertainment, fun or joy. | *“I'm sad to say that the text generators don't have much logic. But it was still very fun to try taking pieces of its responses for my story.”* | 0.857 (6) |
| Idea generation | A student refers to idea(s), student or AI thinking, AI expressions and responses | *"It can help me think of the ideas I don't have."* | 1 (7) |
| Interesting | A student explicitly expresses curiosity, excitement or interest. | *"So interesting"* | 1 (12) |
| Language Issues | A student refers to word(s), sentence(s), grammar or vocabulary, positive or negative, in relation to using AI-generated text | *"I felt so exciting because using AI, when they showed up their words, it's so funny and it would be useful to use them in my short story."* | 1 (11) |
| Novelty | A student expresses the student's first chance, experience, time or unfamiliarity of writing a story using their own words and a machine's words | *"I am not yet familiar with using Al-generated text and for now I still rely on most my brain .however I believe I will grow into it and improve ."* | 1 (7) |
| Own words | A student expresses a preference for any aspect of the student's own words or sentences over a machine's words in story writing. | *"It's more easy just using own words"* | 1 (5) |

Appendix 2.

Coding application

| Student ID | How do you feel writing a short story using your own words and a machine’s words? | AI limitations | Convenience | Fun | Idea generation | Interesting | Language issues | Negative | Novelty | Own words | Positive |
|---|---|---|---|---|---|---|---|---|---|---|---|
| 1 | It can help me think of the ideas I don't have. | | | | 1 | | | | | | 1 |
| 2 | More confident | | | | | | | | | | 1 |
| 3 | So interesting | | | | | 1 | | | | | 1 |
| 4 | Complicated | | | | | | | 1 | | | |
| 5 | Amazing and surprisingly | | | | | | | | | | 1 |

| | | | | | | |
|---|---|---|---|---|---|---|
| 6 | Tired | | | 1 | | |
| 7 | Excited | | | | | 1 |
| 8 | Cool | | | | | 1 |
| 9 | Enjoyable | 1 | | | | 1 |
| 10 | interesting as this is the first time using this method to create a story | | 1 | | 1 | 1 |
| 11 | Interesting, first time ever | | 1 | | 1 | 1 |
| 12 | I feel incredible | | | | | 1 |
| 13 | It something I cannot create as normal, quite good | | | | | 1 |

| | | | | | | | | |
|---|---|---|---|---|---|---|---|---|
| 14 | It’s more easy just using own words | | | 1 | 1 | | 1 | |
| 15 | novel,haven't tried before(as school will not allow us to do so) | | | | | 1 | | |
| 16 | I can generate some interesting idea and help me to write a story | 1 | 1 | | | | | 1 |
| 17 | It is kinda new to me. | | | | | 1 | | |
| 18 | Machine may come up with some astonishing expressions human won't use because we often write in similar | 1 | | | | | | 1 |
| | patterns we are used to, but machine's help us think out of the box. Therefore I think AI writing is pretty advantageous. | | | | | | | |
| 19 | I am not yet familiarwith using Al-generated text and for now I still rely on most my brain .however I believe I will grow into it and improve . | | | | | 1 | 1 | 1 |
| 20 | I'm sad to say that thetext generators | 1 | 1 | 1 | 1 | | | 1 |

| | | | | |
|---|---|---|---|---|
| | don't have much logic. But it was still very fun to try taking pieces of its responses for my story. | | | |
| 21 | AI is so bad | | 1 | |
| 22 | Maybe  good | | | 1 |
| 23 | I fell pretty good,I mean it's amazing to see someone helping me. | | | 1 |
| 24 | I feel so interesting , is a good try. | 1 | | 1 |
| 25 | Good | | | 1 |
| 26 | It good | | | 1 |

| | | | | | | | | | |
|---|---|---|---|---|---|---|---|---|---|
| 27 | It will be very interesting | | | | 1 | | | | 1 |
| 28 | Machine can give me more ideas and it can write some difficult word. | | | 1 | | 1 | | | 1 |
| 29 | Using my own words is a little harder but faster and using a machine might takes some timebut it’s funny. | 1 | 1 | | | 1 | 1 | 1 | 1 |
| 30 | I felt so exciting because using AI, when they showed up their words, it’s so | | 1 | | | 1 | | | 1 |

| | | | | | | | | | | |
|---|---|---|---|---|---|---|---|---|---|---|
| | funny and it would be useful to use them in my short story. | | | | | | | | | |
| 31 | I feel way smarter. | | | | | | | | | 1 |
| 32 | Nice and fun. I can see what the AI thinks and how differently and how interesting the AI can think. It could be weird and funny. But it was such a nice interesting experience and i learnt honestly a lot | | 1 | 1 | 1 | | 1 | 1 | | 1 |
| 33 | I feel more powerful than the ai because the | 1 | | | | 1 | 1 | | 1 | 1 |

| | | | | | |
|---|---|---|---|---|---|
| | ai doesn’t understand simple sentences and makes it too complicated. | | | | |
| 34 | excellent | | | | 1 |
| 35 | 250 | | | | |
| 36 | Get high marks | | | | 1 |
| 37 | no | | | | |
| 38 | I don’t have any feeling | | | | |
| 39 | Amazing | | | | 1 |
| 40 | Joyful. | | 1 | | 1 |
| 41 | It's not bad. It s make writing more easy | 1 | | | 1 |
| 42 | Interesting | | | 1 | 1 |

| | | | | | | |
|---|---|---|---|---|---|---|
| 43 | Good | | | | | 1 |
| 44 | difficult | | | | 1 | |
| 45 | helpful | | | | | 1 |
| 46 | I think it's kinda usefulbut sometimes it doesn't writes the way I wanted it to be | 1 | | | 1 | 1 |
| 47 | Interesting and challenging | | 1 | | 1 | 1 |
| 48 | Some machine's word is i never see it , the words is difficult but it can let my | 1 | 1 | 1 | 1 | 1 |

| | | | |
|---|---|---|---|
| | story more interesting ; if i use my own words maybe i would not always can use the correct grammarbecaus e I 'm not carefully. | | |
| 49 | useful | | 1 |
| 50 | Good | | 1 |
| 51 | Convenience | 1 | 1 |

| | | | | | | | | |
|---|---|---|---|---|---|---|---|---|
| 52 | 崩潰（劇情no sense) | | | | | 1 | | |
| 53 | Sometimes quite confusing but also quite convenient | 1 | | | | 1 | | 1 |
| 54 | Interesting | | | 1 | | | | 1 |
| 55 | Machine know what I want to write and writer no grammar mistakes | | | | 1 | | | 1 |
| 56 | it may be easily to think of ideas or sentences to write | 1 | 1 | | 1 | | | 1 |
| 57 | Quite cool, having the chance to write with | | | | | | 1 | 1 |

| | | | | | | |
|---|---|---|---|---|---|---|
| | AI is a good experience | | | | | |
| 58 | Cool | | | | | 1 |
| 59 | Difficult | | | | 1 | |
| 60 | More convenient and can use less time to write with an AI | 1 | | | | 1 |
| 61 | Happy | | | | | 1 |
| 62 | The machine’s words are more high-class | | | 1 | | 1 |
| 63 | Machine’s words is better than mine | | | 1 | | 1 |
| 64 | Feel good | | | | | 1 |
| 65 | Interesting | | 1 | | | 1 |
| 66 | I like to use my own words more | 1 | 1 | | 1 | |
| 67 | Amazing | | | | | 1 |